\documentstyle[12pt,epsf]{article}

\begin{document}
\newcommand{\be}{\begin{eqnarray}}
\newcommand{\dlq}{\lq\lq}
\newcommand{\ee}{\end{eqnarray}}
\newcommand{\ben}{\begin{eqnarray*}}
\newcommand{\een}{\end{eqnarray*}}
\newcommand{\stackeven}[2]{{{}_{\displaystyle{#1}}\atop\displaystyle{#2}}}
\newcommand{\lsim}{\stackeven{<}{\sim}}
\newcommand{\gsim}{\stackeven{>}{\sim}}
\renewcommand{\baselinestretch}{1.0}
\newcommand{\as}{\alpha_s}
\def\eq#1{{Eq.~(\ref{#1})}}
\def\fig#1{{Fig.~\ref{#1}}}
\begin{flushright} 
NT@UW--01--015 
\end{flushright}
\vskip 20pt
\begin{center}
\begin{title}
\title{\Large\bf Ioffe Time in Double Logarithmic \\[5mm] Approximation}
\end{title}
\vskip 20pt
\begin{author}
\author{Yuri V. Kovchegov $^{1}$  and Mark Strikman $^{2}$}\\ ~~ \\
{\it $^1$ Department of Physics, University of Washington, Box 351560} \\ 
{\it Seattle, WA 98195 } \\ ~~ \\
{\it $^2$ Department of Physics, Pennsylvania State University} \\
{\it University Park, PA 16802} \\ ~~ \\ ~~ \\
\end{author}
\end{center}
\begin{abstract} 
We analyze the light cone (Ioffe) time structure of the gluon
distribution function in the double logarithmic approximation. We show
that due to QCD evolution Ioffe equation is modified. The
characteristic light cone time of the gluons does not increase as fast
with increasing energy (decreasing Bjorken $x$) as predicted by the
parton distributions exhibiting Bjorken scaling due to the increase of
the transverse momenta of the gluons in the DGLAP ladder.
\end{abstract}

\section{Introduction}

It was first observed in Quantum Electrodynamics
\cite{LP} that photon emission in the scattering process of
electrons propagating through a medium occurs over distances
increasing with energies. It their seminal paper \cite{GIP} Gribov,
Ioffe and Pomeranchuk have demonstrated that at high energies large
longitudinal distances, which are now usually referred to as coherent
length distances, $l_c$, become important in the virtual
photon--nucleon scattering when considered in the rest frame of the
target. (One can also formulate this statement in the frame
independent language by using light-cone time along the reaction
axis.) In the following years Ioffe \cite{ioffe} built an explicit
connection between the deep inelastic scattering amplitude and the
space-time representation of the correlator of the electromagnetic
currents. He analyzed the first deep inelastic data from SLAC assuming
the exact Bjorken scaling and demonstrated that the longitudinal
distances in the the Bjorken limit are growing as
\be\label{lcmax}
l_c \propto \frac{1}{m_N \, x},
\ee
where $m_N$ is the target nucleon's mass. A natural implication of the
increase of longitudinal distances was the parallel observation by
Gribov \cite{gribov} that large shadowing effects should be present in
the deep inelastic scattering off nuclei at small enough $x$ when $l_c
\gg 2 R_A$, leading to the saturation behavior of 
$F_{2A} \propto A^{2/3} Q^2 \ln (1/x)$.

Qualitatively one can understand the pattern of \eq{lcmax} by
considering the energy denominator for the transition of the virtual
photon to some excited Fock state consisting of quarks and gluons with
the invariant mass $M^2$. Indeed the uncertainty principle tells us
that the lifetime of this state in the nucleon rest frame is the
inverse of the energy non-conservation in the $\gamma^*\to "M"$
transition and hence $\sim {2q_0 / (M^2+Q^2)}$, which for $M^2$
comparable to $Q^2$ leads to \eq{lcmax}. Bjorken \cite{Bj87} has build
an explicit picture of the space time evolution of the quark-antiquark
pair in the aligned jet model of DIS for the transition $\gamma^*\to
q\bar q $ and demonstrated that even though the pair is produced at
zero transverse separation it reaches hadron size scale over the time
comparable to the coherent length $l_c$ of \eq{lcmax}.

Further studies of $l_c$ were performed in \cite{fs88} in the aligned
jet model. It was demonstrated that in this model the average value of
$M^2$ is $\left<M^2\right> \approx Q^2$, leading to $l_c={1\over
2m_Nx}$.  Similar numbers were obtained in \cite{fs88} by explicit
consideration of the Ioffe representation assuming that the structure
function exhibits Bjorken scaling and that the nucleon's structure
function behaves as $F_{2N}(x)\to const$ in the $x\to 0$ limit.

More recently in a number of papers the coordinate representation of
nucleonic and nuclear correlators of electromagnetic currents was
considered using the current parameterizations of parton densities and
employing the Bjorken scaling approximation for the Ioffe expressions
\cite{PillerWeise,Mankiewicz}. The essential distances in the coherent 
diffractive interactions with nuclei which lead to the leading twist
shadowing were investigated in \cite{fgs} and coherence distances were
found to be of the order of $\sim 1/2m_Nx$ for moderate $Q^2$ but were
decreasing with increasing $Q^2$.

The aim of this paper is to investigate how the scaling violation
which is especially strong in the region of small $x$ affects previous
conclusions for the value of $l_c$.  Qualitatively we can expect that
due to production of intermediate states with a large invariant mass
in the ladder-type kinematics the coherence time $l_c$ should be
substantially reduced. As a first step we will use a double
logarithmic approximation which emphasizes the ladder kinematics. In
Sect. 2 we will consider the double logarithmic evolution with fixed
strong coupling constant and in Sect. 3 we will analyze the case of
running coupling constant in the evolution equations.

In Sect. 4 we will study how the new value for $l_c$ generated by the
double logarithmic evolution affects the onset of non-linear effects
such as saturation \cite{gribov,glrmq,mv,mbal} in a nucleus. To do
that we will have to estimate at what energies the coherence length
becomes larger than the nuclear diameter $2 R_A$. As we will observe
the double logarithmic evolution slows down the onset of these
phenomena with energy.

We want to analyze the structure of the gluon distribution in a proton
in terms of the Ioffe time variable $u = p \cdot x$, where $p_\mu$ is
the target nucleon's momentum and $x_\mu$ is the parton's
coordinate. To avoid confusion throughout the paper we will write $z$
for the Bjorken $x$ variable. The gluon distribution function can be
represented as a Fourier transform in light cone time $u$ as
\cite{ikl}
\be\label{ft}
z G (z, Q^2) \, = \, \int_{-\infty}^{\infty} du \ e^{i u z} \, f (u,
\nu)
\ee
with $\nu = p \cdot q / m_N$ and $Q^2 = - q^2$ the photon's
virtuality. Here the function $f(u,\nu)$ in general also depends on
the transverse momentum squared ${\underline q}^2$. Inverting the
Fourier transform of \eq{ft} we write
\be\label{invf}
f (u, \nu) \, = \, \int_{0}^{\infty} \, \frac{dz}{2 \pi} \, e^{- i u
z} \, z G (z, 2 m_N \nu z) \, - \, \mbox{complex conjugate}.
\ee
Therefore, to determine the coherence length $l_c$, or, equivalently,
the characteristic Ioffe time $u$ of the gluon distribution we have to
find the function $f(u,\nu)$ using the inverse Fourier transform of
\eq{invf} and extract the typical values of $u$ out of it. 

\section{Ioffe Time Evolution of the Double Logarithmic Structure Function: 
Fixed Coupling Case}

We begin by considering the gluon distribution function of a hadron in
the double logarithmic approximation. The DGLAP equation \cite{dglap}
for the gluon structure function in the double logarithmic
approximation is
\be\label{dldglap}
Q^2 \frac{\partial}{\partial Q^2} \ G(z, Q^2) \, = \, \frac{\as
(Q^2)}{2 \pi} \, \int_z^1 \, \frac{dz'}{z'} \, \gamma_{GG}
\left(\frac{z}{z'} \right) \, G (z', Q^2)
\ee
where the gluon--gluon splitting function at small $z$ is given by
\be\label{spl}
\gamma_{GG} (z) \, \approx \, \frac{2 N_c}{z}.
\ee
To solve \eq{dldglap} we for simplicity choose the following initial
condition at some not very large value of photon's virtuality $Q_0^2$
\be\label{init}
G (z, Q_0^2) \, = \, \delta (z - z_0).
\ee
In this section we will consider the ``toy model'' case when the
strong coupling constant is fixed. Then the solution of \eq{dldglap}
is given by
\be\label{fcglue}
z G (z, Q^2) \, = \, \int \ \frac{dn}{2 \pi i} \, \left( \frac{z_0}{z}
\right)^{n-1} \, \left( \frac{Q^2}{Q_0^2} \right)^{\frac{\as N_c}{\pi} 
\frac{1}{n-1}},
\ee
where the integral over $n$ runs along a straight line parallel to
imaginary axis to the right of all the singularities of the
integrand. Using \eq{fcglue} in \eq{invf} we obtain
\ben
f (u, \nu) \, = \, \int_{0}^{\infty} \, \frac{dz}{2 \pi} \, e^{- i u
z} \, \int \ \frac{dn}{2 \pi i} \, \left( \frac{z_0}{z}
\right)^{n-1 - \frac{\as N_c}{\pi} 
\frac{1}{n-1}} \, \left( \frac{\nu}{\nu_0} \right)^{\frac{\as N_c}{\pi} 
\frac{1}{n-1}}
\een
\be\label{fcf1}
- \, \mbox{complex conjugate}.
\ee
Performing the $z$ integration in \eq{fcf1} yields
\ben
f (u, \nu) \, = \, - \frac{1}{|u|} \, \int \ \frac{dn}{2
\pi^2} \, \Gamma \left(2 - n + \frac{\as N_c}{\pi} \frac{1}{n-1}\right) 
\een
\be\label{fcf2}
\times \sin \left[ \frac{\pi}{2} \, 
\left( n - \frac{\as N_c}{\pi} \frac{1}{n-1} \right)  \right] 
(u z_0)^{n-1 - \frac{\as N_c}{\pi} \frac{1}{n-1}} \, \left(
\frac{\nu}{\nu_0} \right)^{\frac{\as N_c}{\pi}
\frac{1}{n-1}}
\ee
where the integration over $n$ can be approximated by the saddle point
method around
\be
n_0 - 1 \, = \, \sqrt{\frac{\as N_c}{\pi} \, \ln \frac{\nu}{u
z_0 \nu_0}\frac{1}{\ln u z_0}}
\ee
to give
\ben
f (u, \nu) \, = \, - i \frac{1}{|u|} \, \left( \frac{\as
N_c \pi \ln \frac{\nu}{u z_0 \nu_0}}{\ln^3 u z_0} \right)^{1/4} 
\een
\be\label{fcf3}
\times \, \exp
\left( 2 \sqrt{\frac{\as N_c}{\pi} \, \ln \frac{\nu}{u z_0 \nu_0} 
\ln u z_0}\right).
\ee
\eq{fcf3} gives the Fourier image in Ioffe time of the gluon structure 
function in a proton. Our goal is to determine which values of the
variable $u$ dominate in \eq{ft} with the function $f (u,\nu)$ given
by \eq{fcf3}. Most of the $z$ and $\nu$ dependence in \eq{fcf3} is
driven by the exponential function in it. To determine the
characteristic values of $u$ we have to find which $u$ gives the
maximum of the function in the exponent. One can readily see that the
maximum is reached at
\be\label{fcu}
u^*_{fc} \, = \, \frac{1}{z_0} \, \sqrt{\frac{\nu}{\nu_0}}.
\ee
Another way of determining the typical value of $u$ is by finding the
median of the integral over $u$ in \eq{ft} \cite{fs88}. Our numerical
estimates showed that the discrepancy between that method and the one
presented above is not significant in the double logarithmic
approximation and thus \eq{fcu} does give a good estimate of the
typical $u$ in the gluon distribution function.

One can see that the gluon smearing of \eq{fcu} is not as large as one
would naively expect from the Ioffe equation where in the Bjorken
scaling limit there assumed to be no $\nu$ dependence in the integrand
of \eq{ft}. There the spread of partons is limited only by the Fourier
exponent in \eq{ft} yielding the maximum possible value of $u$ to be
\be\label{lim}
u^*_{max} \, = \, \frac{1}{z}
\ee
which translates itself into the well-known coherence length of the
gluon in the proton's rest frame $l_{coh} = 1 / 2 m_N z$. Recalling
that $\nu = Q^2 / 2 m_N z$ one can see that the time in \eq{fcu} grows
much slower with decreasing $z$ than the time given by \eq{lim}.

To understand what happens it is instructive to rewrite \eq{fcu} in
terms of the $x_-$ coordinate and $q_-$ momentum of the last gluon in
the ladder. Since we are working in the infinite momentum frame where
$p_\mu = (p_+, 0, {\underline 0})$ we write $u = p \cdot x = p_+ x_-$
and $\nu = p \cdot q / m_N = p_+ q_- / m_N$ which leads us to the
following expression for the light cone lifetime of the last gluon in
the ladder
\be\label{fcx}
x^*_{fc-} \, = \, \frac{2 \, q_-}{\sqrt{Q_0^2 \, z_0 \, 2 \, p_+ \,
q_-}} \, \approx \, \frac{2 \, q_-}{\sqrt{Q_0^2 \, z_0 \, s}},
\ee
where we introduced the center of mass energy $s = (p+q)^2 \approx 2
p_+ q_-$.  On the other hand in a frame where several evolution rungs
start in the probe \cite{mbal} we can write down the light cone
lifetime of a (longitudinally) soft gluon as
\be\label{lc}
x^*_- \, = \, \frac{2 \, q_-}{{\underline q}^2}.
\ee
To estimate the typical ${\underline q}^2$ of a gluon in the DLA DGLAP
evolution taken at a given fixed value of $q_-$ (or, equivalently,
$s$) we first note that in the saddle point approximation \eq{fcglue}
leads to
\be\label{dlaxg}
z G (z, Q^2) \, \propto \, \exp
\left( 2 \sqrt{\frac{\as N_c}{\pi} \, \ln \frac{Q^2}{Q^2_0} \, 
\ln \frac{z_0}{z}}\right).
\ee
Substituting $Q^2 \approx {\underline q}^2$ and $z \approx {\underline
q}^2 / 2 p_+ q_-$ we obtain the typical transverse momentum squared of
the gluons
\be\label{trfc}
\left< {\underline q}^2 \right>_{fc} \, \approx \, 
\sqrt{Q_0^2 \, z_0 \, 2 \, p_+ \, q_-} \, \approx \, \sqrt{Q_0^2 \, z_0 \, s},
\ee
which, after being substituted into \eq{lc} gives us \eq{fcx}. Now the
difference between Ioffe times in the distribution function exhibiting
Bjorken scaling and the double logarithmic gluon distribution becomes
apparent. As we increase the center of mass energy $s$ by increasing
$q_-$ the typical transverse momentum of the distribution function
without any QCD evolution would remain constant, while in the case of
double logarithmic evolution it would increase with energy (see
\eq{trfc}) leading to shorter light cone lifetimes in this latter
case.  The diffusion of the typical transverse momentum with energy is
so fast in our \eq{trfc} because in this section we considered the toy
model of fixed coupling DLA DGLAP equation.

\section{Ioffe Time Evolution of the Double Logarithmic Structure Function: 
Running Coupling Case}

Let us repeat the calculation of the previous chapter for the running
coupling case. The running coupling constant in \eq{dldglap} will be
taken below at the one-loop level
\be
\as (Q^2) \, = \, \frac{1}{b \ln Q^2/\Lambda^2}
\ee
with the beta function $b = \frac{11 N_c - 2 N_f}{12 \pi}$.

The solution of the double logarithmic DGLAP equation for gluon
distribution function (\ref{dldglap}) with the splitting function of
\eq{spl} and the initial conditions given by \eq{init} is
\be\label{dlglue}
z G (z, Q^2) \, = \, \int \ \frac{dn}{2 \pi i} \, \left( \frac{z_0}{z}
\right)^{n-1} \, \left( \frac{\ln (Q^2/\Lambda^2)}{\ln (Q_0^2/\Lambda^2)} 
\right)^{\frac{N_c}{\pi b (n-1)}},
\ee
where, just like in \eq{fcglue} the integral over $n$ runs along a
straight line parallel to imaginary axis to the right of all the
singularities of the integrand.  Substituting \eq{dlglue} into
\eq{invf} we write for the Ioffe time representation of the gluon
structure function \cite{ioffe}
\ben
f (u, \nu) = \int_{0}^{\infty} \, \frac{dz}{2 \pi} \, e^{- i u z} \,
\int \ \frac{dn}{2 \pi i} \, \left( \frac{z_0}{z}
\right)^{n-1} \, \left( \frac{\ln (2 \nu m_N z/\Lambda^2)}{\ln 
(2 \nu_0 m_N z_0/\Lambda^2)} 
\right)^{\frac{N_c}{\pi b (n-1)}} \, - 
\een
\be\label{f1}
- \, \mbox{complex conjugate}.
\ee
Integration over $z$ in \eq{f1} is rather complicated but can be
simplified if with the leading logarithmic accuracy we substitute $z$
by $1/u$ in the ratio of the logarithms in it. The rest of the
integral can be done easily yielding
\ben
f (u, \nu) \, = \, - \frac{1}{|u|} \int \ \frac{dn}{2
\pi^2} \, \Gamma (2-n) \, \sin \left( \frac{\pi n}{2}\right) \, 
\een
\be\label{f2}
\times (u z_0)^{n-1} \,   
\left( \frac{\ln (2 \nu m_N / u \Lambda^2)}{\ln (2 \nu_0 m_N z_0/\Lambda^2)} 
\right)^{\frac{N_c}{\pi b (n-1)}} .
\ee
The expression in \eq{f2} can be evaluated by the saddle point
method. The position of the saddle point is given by
\be
n_0 - 1 \, = \, \sqrt{\frac{N_c}{\pi b} \, \ln \left( \frac{\ln (2 \nu m_N /u
\Lambda^2)}{\ln (2 \nu_0 m_N z_0/\Lambda^2)} \right) \, \frac{1}{\ln (u z_0) }}
\ee
and \eq{f2} becomes
\ben
f (u, \nu) \, = \, - i \, \frac{1}{2 \pi^2 |u|} \, \left[
\frac{\frac{N_c \pi}{b} \ln \left( \frac{\ln (2 \nu m_N /u
\Lambda^2)}{\ln (2 \nu_0 m_N z_0/\Lambda^2)} \right)}{\ln^3 (u z_0)} \right]^{1/4} 
\een
\be\label{f3}
\times \, \exp \left( 2 \sqrt{\frac{N_c}{\pi b} \, \ln \left( \frac{\ln 
(2 \nu m_N /u \Lambda^2)}{\ln (2 \nu_0 m_N z_0/\Lambda^2)} \right) \,
\ln (u z_0) } \right).
\ee
\eq{f3} yields the distribution in light cone Ioffe time of the gluon 
structure function in a proton. Similarly to the previous section in
order to find the characteristic value of $u$ we need to find the
maximum of the function in the power of the exponent in
\eq{f3}. Defining $L = \ln (2 \nu m_N z_0/\Lambda^2)$ and $L_0 = \ln (2
\nu_0 m_N z_0/\Lambda^2)$ together with $\zeta = \ln (u z_0) / L$ allows
us to rewrite the function under the square root in the exponent of
\eq{f3} as
\be\label{sq}
h (\zeta) = \zeta L \ln \frac{L (1 - \zeta)}{L_0}.
\ee
For $L \gg L_0$ the maximum of \eq{sq} is reached at the value of
$\zeta$ that can be very well approximated by the formula
\be\label{max}
\zeta^* \, \approx \, 1 - \frac{1}{\ln L/L_0}
\ee
which is found from the usual extremum condition
\be
h' (\zeta^*) = \ln \frac{L}{L_0} + \ln ( 1 - \zeta^*) - \frac{\zeta^*}{1 -
\zeta^*} \approx \ln \frac{L}{L_0} - \frac{\zeta^* (2 - \zeta^*)}{1 -
\zeta^*} = 0.
\ee
\eq{max} can be rewritten in terms of the Ioffe time and $\nu$ as
\be\label{imax}
u^* \, \approx \, \frac{1}{z_0} \, \left( \frac{2 \, \nu \, m_N \,
z_0}{\Lambda^2} \right)^{1 - \ln^{-1} \left( \frac{\ln (2 \nu m_N z_0/
\Lambda^2)}{\ln (2 \nu_0 m_N z_0/\Lambda^2)} \right)}
\ee
or in terms of $x_-$ and $q_-$ as
\ben
x^*_{rc-} \, = \, \frac{2 \, q_-}{\Lambda^2} \, \left( \frac{2 \, p_+
\, q_- \, z_0}{\Lambda^2} \right)^{- \ln^{-1} \left( \frac{\ln (2 p_+
q_- z_0 / \Lambda^2)}{\ln (Q_0^2/\Lambda^2)} \right)} 
\een
\be\label{rcx}
\approx \, \frac{2 \, q_-}{\Lambda^2} \, 
\left( \frac{s \, z_0}{\Lambda^2} \right)^{ - \ln^{-1}
\left( \frac{\ln (s z_0 / \Lambda^2)}{\ln (Q_0^2/\Lambda^2)} \right)}.
\ee
To understand this result we, similarly to the previous section first
note that in the saddle point approximation the gluon distribution of
\eq{dlglue} is proportional to
\be\label{rcgl}
z G(z, Q^2) \, \propto \, \exp \left( 2 \sqrt{\frac{N_c}{\pi b} \, \ln
\left( \frac{\ln (Q^2 / \Lambda^2)}{\ln (Q^2_0/\Lambda^2)} \right) \,
\ln \frac{z_0}{z} } \right)
\ee
with the typical transverse momentum for fixed $q_-$ given by
\be\label{drc}
\left< {\underline q}^2 \right>_{rc} \, \approx \, \Lambda^2 \, 
\left( \frac{2 \, p_+ \, q_- \, z_0}{\Lambda^2} \right)^{\ln^{-1} 
\left( \frac{\ln (2 p_+ q_- z_0 / \Lambda^2)}{\ln (Q_0^2/\Lambda^2)} \right)} 
\, \approx \, \Lambda^2 \, \left( \frac{s \, z_0}{\Lambda^2} \right)^{\ln^{-1} 
\left( \frac{\ln (s z_0 / \Lambda^2)}{\ln (Q_0^2/\Lambda^2)} \right)}.
\ee
Substituting \eq{drc} into \eq{lc} readily yields us
\eq{rcx}. Thus again the DGLAP evolution makes transverse momenta in 
the gluon distribution diffuse towards larger values slowing down the
spreading of the gluons in the longitudinal $x_-$ direction as
compared to the standard estimate of \eq{lim}. The growth of the
transverse momenta with $s$ described by \eq{drc} is slower than any
positive power of $s$. Thus it is slower than $\left< {\underline q}^2
\right>_{fc} \, \sim \, \sqrt{s}$ of the fixed coupling case (see
\eq{trfc}). Slower growth of the transverse momentum in the running
coupling case leads to longer light cone coherence times for the
gluons than in the fixed coupling case, as follows from
\eq{lc}. Nevertheless the coherence time is still shorter than the
maximum limit of \eq{lim}.

\section{Discussion}

To illustrate our results we are going to plot the gluon coherence
lengths produced by different evolution scenarios discussed above in
the rest frame of the proton as functions of center of mass energy
$s$. By doing so we would also address the question posed in the
Introduction: at which values of $z$ does the DGLAP evolved gluon
distribution of a nucleon in a nucleus reach the lengths comparable to
the nuclear diameter, so that the non-linear effects involving
multiple rescatterings and mergers of gluon ladders would begin taking
place?

In order to plot the coherence length of the gluons in the rest frame
of the target proton or nucleus we use the simple relationship $u =
1/2 m_N l_{coh}$ and employ Eqs. (\ref{fcx}) and (\ref{rcx}) for the
fixed and running coupling cases correspondingly. We use $\Lambda =
0.3 \, \mbox{GeV}$ and $z_0 = 0.1$. In order for the diffusion of
transverse momenta to start at the same initial value we have to
choose different values of the initial virtuality $Q_0$ for the fixed
and running coupling cases: we use $Q_0 = 4 \, \mbox{GeV}$ in \eq{fcx}
and $Q_0 = 0.8 \, \mbox{GeV}$ in \eq{rcx}. By doing so we make the
initial transverse momentum squared of both evolutions equal to
$q_{init}^2 \approx 20 \, \mbox{GeV}^2$ at $s = 200 \, \mbox{GeV}^2$.

\begin{figure}[t]
\begin{center}
\epsfxsize=13.5cm
\leavevmode
\hbox{ \epsffile{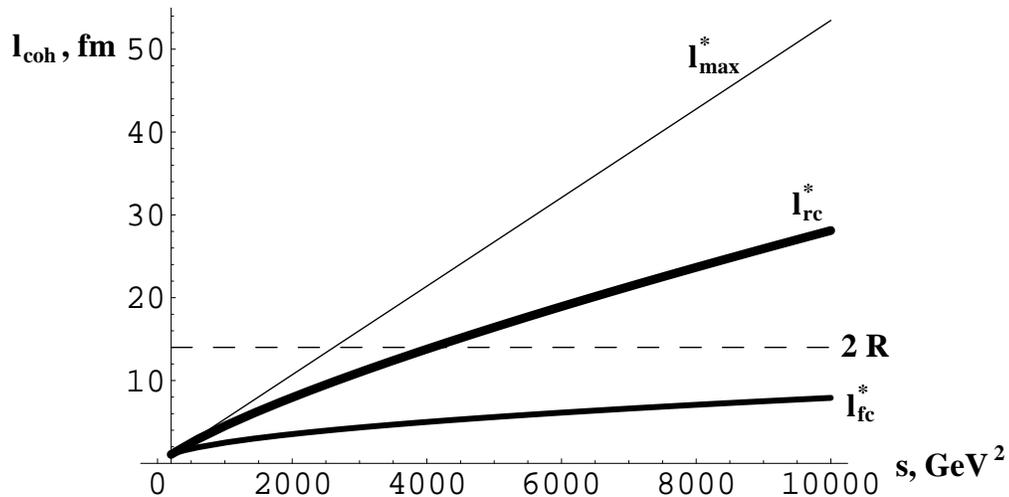}}
\end{center}
\caption{Distances characterizing the spread of gluons in the rest 
frame of the proton. Horizontal line denotes the diameter of some
large nucleus which is taken here to be $14$ fm. The top solid line
corresponds to the upper limit of the \eq{lim}, the lower
(medium-thick) line corresponds to the fixed coupling DLA DGLAP
evolution of \eq{fcx} and the middle (thick) line depicts the running
coupling evolution of \eq{rcx}. }
\label{len}
\end{figure}

The plot of $l_{coh}$ as a function of the center of mass energy $s$
is depicted in \fig{len}. The medium-thick line corresponds to the
fixed coupling DLA DGLAP evolution ($l_{fc}^*$) while the thick line
corresponds to the running coupling case ($l_{rc}^*$). The top solid
line depicts the upper limit on the coherence length given by $l_{coh}
= 1/2 m_N z$, where $z = q_{init}^2 / s$ to insure that this length is
really maximum possible. Finally the horizontal line corresponds to
the diameter of some sample large nucleus taken here to be equal to $2
R = 14 \, \mbox{fm}$. One can see that due to DGLAP evolution the
coherence lengths of gluons do not grow as fast with energy $s$ as
predicted by the well-known estimate of $1/2 m_N z$ (upper line). For
the case of running coupling the coherence length is roughly a half of
the maximum throughout the whole region of $s$
considered. Appropriately, in the case of a nuclear target, the
coherence length of the partons in each individual nucleon reaches the
nuclear diameter at higher values of $s$ than expected from
\eq{lim}. In the fixed coupling case the nuclear diameter is reached
at the energies beyond those shown in \fig{len}, at about $s \approx
30000 \, \mbox{GeV}^2$. In a more realistic case of running coupling
DGLAP evolution the crossover happens slightly above the one predicted
by \eq{lim}, leading to about $30$\% higher $s$ required for the
gluon's coherence length to become comparable to the nuclear radius
(see \fig{len}). To summarize, based on the example of the double
logarithmic approximation we conclude that DGLAP evolution towards
higher $Q^2$ slows down the onset of non-linear effects in the nuclear
wave function.

From the picture presented in \fig{len} one concludes that much
smaller than naively expected values of $z$ (for the same large $Q^2$)
are needed in the nuclei for the parton distributions of the
individual nucleons to start overlapping making non-linear effect
related to nuclear shadowing and parton saturation possible. However
one must keep in mind that the spread of the gluon distribution in
\fig{len} is generated by the DLA DGLAP evolution, that is along with
going towards smaller values of $z$ the evolution also moves towards
larger values of transverse momentum ${\underline q}^2$ as could be
seen from Eqs. (\ref{trfc}) and (\ref{drc}). Going towards smaller $z$
tends to increase light cone times and corresponding coherence lengths
in the target's rest frame. At the same time increasing ${\underline
q}^2$ tends to decrease light cone times, moving the evolution away
from the saturation region of moderate ${\underline q}^2 \sim
Q_s^2$. Thus even though DGLAP evolution seems to slow down the growth
of the light cone times as functions of $z$ as could be seen from
Eqs. (\ref{fcx}) and (\ref{rcx}), this is entirely due to the fact
that DLA DGLAP evolution moves the distributions toward larger
${\underline q}^2$ pushing it away from the saturation region and
non-linear effects.

Of course the gluon distribution generated by the double logarithmic
evolution equation describes the data well only in a rather narrow
kinematic region of small $x$ and large $Q^2$. Therefore our analysis
of light cone Ioffe time structure of the DLA gluon distribution is
also limited to this kinematic region. Further studies involving more
realistic parton densities beyond the double log approximation are
necessary in order to fully quantify the effect of scaling violations
on the light cone structure of distribution functions.

\section*{Acknowledgments}

The authors are very much indebted to Lonya Frankfurt for several
insightful discussions. M.S. thanks the Department of Energy's
Institute for Nuclear Theory at the University of Washington for its
hospitality and the Department of Energy for partial support during
the time this study started. The work of Yu.K. was supported in part
by the U.S. Department of Energy under Grant
No. DE-FG03-97ER41014. The research of M.S. was sponsored in part by
the U.S. Department of Energy under Grant No. DE-FG02-93ER40771.

\end{document}